\documentclass[%
 aps,
 prl,
 twocolumn,
 superscriptaddress,
 showpacs,
 floatfix,
 nofootinbib,
 amsmath,
 amssymb,
 10pt
]{revtex4-2}
\usepackage{graphicx}
\usepackage{hyperref}
\usepackage{xcolor}
\usepackage{natbib}
\usepackage{float}

\newcommand{\mean}[1]{\left\langle #1 \right\rangle}

\newcommand{\meansq}[1]{\left\langle ( #1 )^2\right\rangle}

\def\TRENTo{{\sc t\kern-.05em \lower.5ex\hbox{r}\kern-.025em e\kern-.05em n\kern-.05em t\kern-.09em}o}
\def\iccing{{\sc i\kern-.05em c\kern-.05em c\kern-.05em i\kern-.05em n\kern-.05em g\kern-.05em}}
\def\ccake{{\sc c\kern-.05em c\kern-.05em a\kern-.05em k\kern-.05em e\kern-.05em}}
\def\vUSPhydro{v-{\sc u\kern-.05em s\kern-.05em p\kern-.05em}hydro}
\graphicspath{ {figures/} }

\def\l{\left}
\def\r{\right}

\begin{document}

\title{Unlocking ``imprints" of conserved charges in the initial state of heavy-ion collisions}

\author{Fernando G. Gardim}
\affiliation{Instituto de Ci\^encia e Tecnologia, Universidade Federal de Alfenas, 37715-400 Po\c cos de Caldas, MG, Brazil}
\affiliation{Illinois Center for Advanced Studies of the Universe, Department of Physics, University of Illinois at Urbana-Champaign, Urbana, IL 61801, USA}
\author{Dekrayat Almaalol}
\affiliation{Illinois Center for Advanced Studies of the Universe, Department of Physics, University of Illinois at Urbana-Champaign, Urbana, IL 61801, USA}
\author{Jordi Salinas~San~Mart\'in}
\affiliation{Illinois Center for Advanced Studies of the Universe, Department of Physics, University of Illinois at Urbana-Champaign, Urbana, IL 61801, USA}
\author{Christopher Plumberg}
\affiliation{Natural Science Division, Pepperdine University, Malibu, CA 90263, USA}
\author{Jacquelyn Noronha-Hostler}
\affiliation{Illinois Center for Advanced Studies of the Universe, Department of Physics, University of Illinois at Urbana-Champaign, Urbana, IL 61801, USA}

\date{\today}

\begin{abstract}
Hydrodynamic approaches to modeling relativistic high-energy heavy-ion collisions are based on the conservation of energy and momentum.
However, the medium formed in these collisions also carries additional conserved quantities, including baryon number (B), strangeness (S), and electric charge (Q).
In this Letter, we propose a new set of anisotropic flow observables designed to be exclusively sensitive to the effects of conserved BSQ charge fluctuations, providing insight into the initial state.
Using the recently developed hydrodynamic framework \iccing{}+\ccake{}, we show that these new observables provide a measurable effect  of initial BSQ charge fluctuations (ranging up to $\sim $10\%), which can be tested by experiments.

\end{abstract}

\maketitle

Since the early 2000s the Quark Gluon Plasma (QGP) has been studied using relativistic heavy-ion collisions at both the Large Hadron Collider (LHC) and the Relativistic Heavy-Ion Collider (RHIC). 
Detailed comparisons of theoretical predictions with experimental measurements have demonstrated that the QGP acts as a nearly perfect fluid with the smallest shear viscosity to entropy density ratio ever discovered \cite{Busza:2018rrf, Heinz:2005zg, Moreland:2018gsh, Bernhard:2019bmu, JETSCAPE:2020mzn}.
Significant developments in the field of relativistic viscous hydrodynamics have led to additional insights into the QGP and its properties, including the criteria for QGP formation \cite{Nagle:2018nvi, Summerfield:2021oex, Kanakubo:2022jju, Gardim:2022yds}, the QGP's vorticity \cite{Becattini:2020ngo, Kharzeev:2024zzm}, magnetic fields formed in the QGP \cite{Inghirami:2019mkc}, how jets couple to the QGP \cite{Andrade:2014swa,Tachibana:2017syd}, whether the QGP may manifest critical behavior \cite{An:2021wof}, and the dynamics of the QGP with multiple conserved charges \cite{Karpenko:2013wva,Du:2019obx, Plumberg:2023vkw, Plumberg:2024leb,Monnai:2024pvy,Garcia-Montero:2023gex}.

In these studies, the basic assumption has been that the initial state after the collision is either a state of saturated gluons or is dominated by nucleonic collisions. 
Some recent studies have also considered the possible influence nucleonic substructure on the geometry of the initial state \cite{Petersen:2010zt,Schenke:2014zha,Noronha-Hostler:2015coa,Mazeliauskas:2015vea,Gardim:2017ruc} and have found a preference for some substructure, although they could not constrain the number of hot spots \cite{Moreland:2018gsh}.
So far, however, none of these studies has taken into account the flavor and color carried by the quarks appearing at the sub-nucleonic level.
Only with the recent development of  \iccing{} \cite{Carzon:2019qja} has a code taken into account gluons splitting into quark-antiquark pairs ($g\to q\bar{q}$) that tracked the flavor of the produced quarks. 
These splitting probabilities can be calculated directly within the color glass condensate framework \cite{Martinez:2018ygo}.
Since each quark carries three conserved charges (baryon number B, strangeness S, electric charge Q), an entire framework is required which can propagate BSQ conserved charges through hydrodynamics and into hadronic interactions. 
Recently, a BSQ heavy-ion simulator called \ccake{} \cite{Plumberg:2024leb} was developed which couples to \iccing{} and subsequently evolves the full initial state, including both energy density and BSQ densities. That study indicated that using two particles-of-interest for measurements of identified particle (PID) collective flow could reveal signatures of the $q\bar{q}$ pairs formed in the initial state. Recent studies \cite{Holtermann:2023vwr,Holtermann:2024vdw} have already proposed a wealth of new potential observables that can be considered with one or two particles-of-interest. These observables open up a completely new set of tools for probing and constraining the properties of the QGP. 
However, the range of possible experimental observables which are sensitive to BSQ fluctuations in the initial state has yet to be fully explored.

In this Letter, we propose a unique set of flow observables which, by design, are sensitive exclusively to BSQ fluctuations (i.e., they are identically 1 or 0 in the absence of BSQ fluctuations).
We use the framework of \iccing{}+\ccake{} to make predictions for these observables calculated in Pb-Pb collisions at $\sqrt{s_{NN}} = 5.02$ TeV.
We find that these new experimental flow observables exhibit up to $\sim$10\% effects from the inclusion of BSQ fluctuations in the initial state. 
These observables should be measurable in the future high luminosity runs at the LHC which will provide sufficient statistics to render flow observables with more than just a single particle-of-interest possible.

{\it BSQ Physics.}––Our goal is to use a framework that incorporates the splitting of gluons into $q\bar{q}$ pairs into the initial state where the resulting conserved charges are propagated into the final state with relativistic viscous hydrodynamics, followed by hadronic interactions.
We use event-by-event \TRENTo{} \cite{Moreland:2014oya} initial conditions to generate the initial state energy density, $\varepsilon(\vec{r})$. 
We assume that the resulting $\varepsilon(\vec{r})$  is composed entirely of gluons, allowing us to use it as input for \iccing{} \cite{Martinez:2019jbu, Carzon:2019qja}.
The output from \iccing{} then supplies the initial conditions for the subsequent hydrodynamic evolution and contains fluctuating distributions in both $\varepsilon$ and the conserved charge densities (in baryon number, strangeness, and electric charge), where the latter are constrained to be zero on average.

After the initial conditions have been generated, we feed them into the 2+1 relativistic viscous hydrodynamic code \ccake{} that propagates both $\varepsilon$ and BSQ densities simultaneously assuming no BSQ diffusion.
For the 4-dimension equation of state (EoS) required to close the hydrodynamic equations of motion, we primarily reply upon the lattice QCD EoS from \cite{Noronha-Hostler:2019ayj} which is coupled to the PDG2016+ list \cite{Alba:2017mqu}.
For those fluid cells which do not fit within the range of validity of the Taylor series from lattice QCD, we employ the back up EoS from \cite{Plumberg:2024leb}.
Following the hydrodynamic phase, we freeze-out at a fixed energy density $\varepsilon_\text{FO} =$ 0.276 GeV/fm$^3$ (based on a $T_\text{FO}=150$ MeV at vanishing densities) into particles and allow for hadron decays using an adapted version of \cite{Sollfrank:1990qz} before performing the flow analysis on PID.

We employ the same \TRENTo, \iccing, and \ccake{} parameters as described in \cite{Plumberg:2024leb} that provided a reasonable description of the standard, event-by-event averaged observables such as $dN/dy$, $\l< p_T \r>$, and $v_n\lbrace 2 \rbrace$.
While one could potentially vary the fit parameters further using a Bayesian analysis, in this work we are interested in isolating the effects of BSQ fluctuations, which we anticipate should have a minimal effect on these standard observables.
We ran 6,000 events between $0$--$60\%$ centrality using the \TRENTo{}+\iccing{}+\ccake{} for Pb-Pb at $\sqrt{s_{NN}}=5.02$ TeV.  We calculated PID anisotropic flow for all ground-state baryons and mesons. 
Since the only available LHC PID data  are from the ALICE  \cite{ALICE:2018yph} with $0.2<p_T<3.0$ GeV/$c$, we used the same kinematic cuts.

{\it Anisotropic flow.}––The initial state of heavy-ion collisions has a predominantly elliptical shape, due to the collision dynamics of two Lorentz contracted nuclei.  Sub-leading fluctuations at the nucleonic level then generate higher order azimuthal harmonics (e.g., triangular, quadrangular, etc.).  Because of the nearly perfect fluid nature of the QGP, these harmonics appear in the final hadron transverse momentum spectra $p_T$ via a Fourier decomposition, where their magnitudes are determined by the resulting Fourier coefficients.
The flow vector for a single event is
\begin{equation}
    V_n \equiv v_n e^{in\Psi_n} = \frac{1}{M}\sum_j e^{in\phi_j},
    \label{eq:flow_vec}
\end{equation}
which depends on the magnitude of the flow harmonics $v_n$ and their corresponding event plane angle $\Psi_n$, corresponding to correlating individual particles $j$ with their angles $\phi_j$ and normalized by the total multiplicity $M$ of the event.
Typically these $V_n$ are extracted from the spectra of all charged particles within a fixed kinematic range across $p_T$ and rapidity. 
However, one can also extract $V_n$ from a subset of the measured PID that might be of interest (e.g., kaons or even just $K^+$) , which we refer to as a particle-of-interest (POI).
A POI may also be a particle at a specific $p_T$ or rapidity (see \cite{Borghini:2001vi,Borghini:2000sa,Bilandzic:2013kga,Bilandzic:2010jr,Betz:2016ayq,Moravcova:2020wnf,Bilandzic:2021rgb,Holtermann:2023vwr,Holtermann:2024vdw} and measurements \cite{CMS:2017xgk,CMS:2021qqk,ALICE:2022dtx,ALICE:2022zks,ATLAS:2015qwl,ALICE:2018gif}) but we do not consider such a possibility in this paper. 
As a general notation, we use the superscript $+$ (as in $V_n^+$) to denote a particle (e.g., $\Lambda$ or $K^+$) and the superscript $-$ (as in $V_n^-$) to indicate its corresponding antiparticle (e.g., $\bar{\Lambda}$ or $K^-$). 
To be clear, the index $\pm$ does not necessarily correspond to the \emph{electric} charge of the particle itself (as in the example of $\Lambda$ above). 
Then we use $h$ to indicate that both the particle and its antiparticle, i.e., $h=\Lambda+\bar{\Lambda}$ or $h=K^++K^-$ are combined together in the analysis.
For $\Lambda$'s specifically it is not possible experimentally to distinguish them from $\Sigma^0$'s. 
Therefore, we define the $+$ as $\Lambda+\Sigma^0$, the $-$ as $\bar{\Lambda}+\bar{\Sigma}^0$, and $h=\Lambda+\Sigma^0+\bar{\Lambda}+\bar{\Sigma}^0$ includes all 4 particles. 
\begin{figure}[h!]
    \centering
    \includegraphics[width=0.5\textwidth]{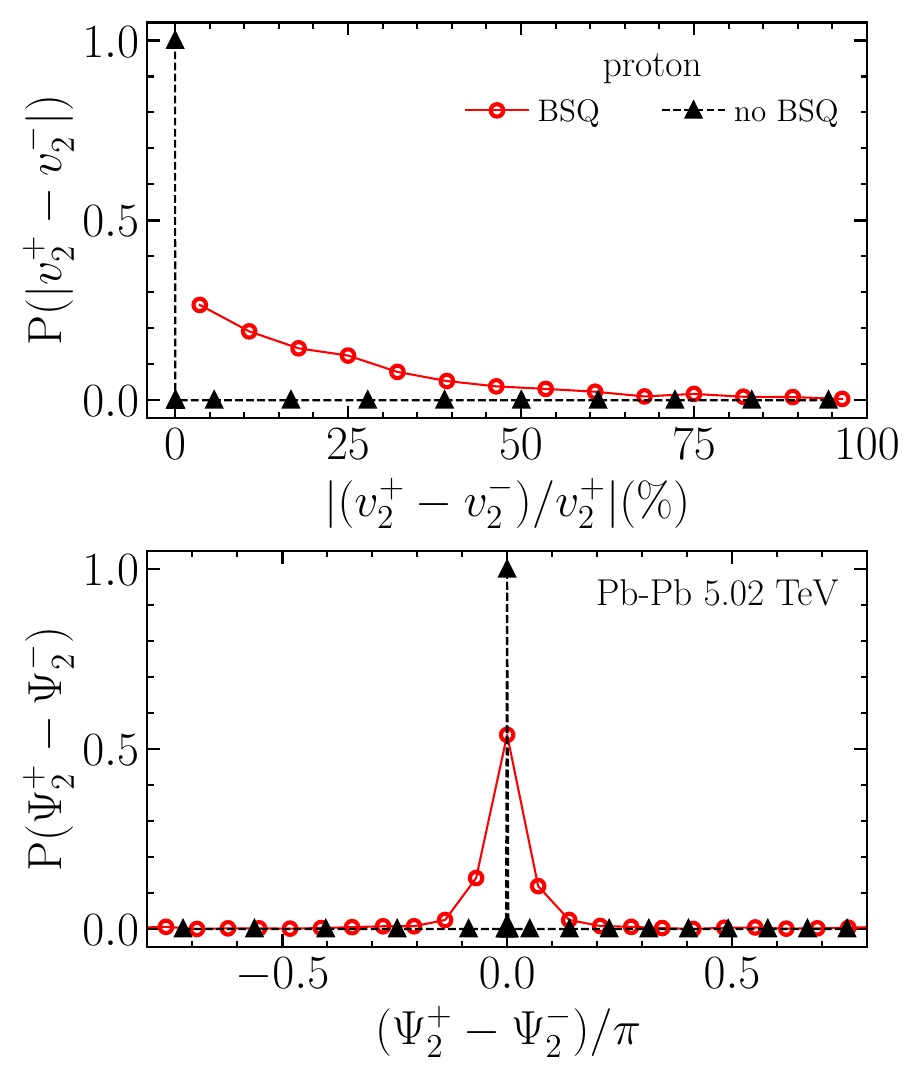}
    \caption{Probability of finding differences between protons and antiprotons for the anisotropic flow (top) and their event-plane angles (bottom) for Pb-Pb collisions at $\sqrt{s_{NN}}=5.02$ TeV and 0-5\% centrality. Results for no conserved charges (\ccake{}) are represented by black triangles, and with BSQ fluctuations (\iccing{}+\ccake{})  in red circles.}
    \label{fig:BSQ_effect}
\end{figure}

In the absence of BSQ charge fluctuations, the flow vector for a particle species should be identical\footnote{If one does Monte Carlo sampling of the spectra to produce hadrons to be fed into a hadron transport code, some very tiny deviation may occur due to finite statistical effects, although this is expected to be negligible in large systems.  Furthermore, Monte Carlo sampling of hadrons+BSQ fluctuations would in general only lead to even larger fluctuations and thus would enhance the observables shown here.} to that of its antiparticle, such that $v_n^+=v_n^-$ and $\Psi_n^+=\Psi_n^-$. 
Fig. \ref{fig:BSQ_effect} illustrates the distribution of $|v_n^+-v_n^-|$ (top) and $\Psi_n^+-\Psi_n^-$ (bottom) for $p$ versus $\bar{p}$.
The results show clearly that in the absence of BSQ fluctuations (\ccake{} only), $V_n^+=V_n^-$ on an event-by-event basis due to the delta function at 0.
However, once we allow for initial state BSQ fluctuations generated and evolved by \iccing{}+\ccake{}, we obtain non-trivial distributions for $|v_n^+-v_n^-|$ and $(\Psi_n^+-\Psi_n^-)$, observing variations of up to 50\% in the flow with BSQ fluctuations. 
Thus, $V_n^+\neq V_n^-$ for \iccing{}+\ccake{} and in order to construct observables sensitive to BSQ fluctuations, we must use POI from PID.

Another important point to highlight from Fig.\ \ref{fig:BSQ_effect} is that both distributions are centered around zero.
The mean of each distribution is 0 because in our simulations we have no \textit{net} BSQ densities (only fluctuations around 0).
Thus, we do not expect a \emph{global} imbalance of $\langle v_n^+\rangle \neq \langle v_n^-\rangle$ (where $\langle \cdots\rangle$ indicates averaging over events).
This last statement will no longer hold in general when one considers lower beam energies.

It was shown in \cite{Plumberg:2024leb} that all charged particle $v_n\left\{2\right\}$ exhibited no significant effect from BSQ fluctuations, which confirms past findings of the nearly linear response of energy density ($\varepsilon$) eccentricities to all charge particle flow \cite{Teaney:2010vd,Qin:2010pf,Gardim:2011xv,Qiu:2011iv,Gardim:2014tya}. 
Thus, we can define the PID flow vectors  as a contribution predominately from energy $V^{\varepsilon}_n$  along with a small perturbation induced by the gluon splitting $\delta V_n^{\pm}$: 
\begin{equation}
    V_n^\pm  = V_n^{\varepsilon} + \delta V_n^\pm.
    \label{eq:flow_id}
\end{equation}
Experimentally, the anisotropic flow must be rotationally invariant such that one uses 2-particle correlations, namely,
\begin{equation}
   v_n^{\rm ch}\{2\}\equiv \sqrt{\mean{\left(v_n^{\rm ch}\right)^2}} = \sqrt{\mean{|V_n^{\rm ch}|^2}}.
   \label{eq:vn2}
\end{equation}
The average is often calculated by considering all charged particles $v_n^{\rm ch}$. 
In the case of PID, so far only observables using one POI (1POI) have been measured experimentally, meaning that one typically correlates a PID with both its particle and antiparticle (e.g., $h\equiv p+\bar{p}$ as was done in \cite{ALICE:2018yph}). 
We can then define the averaged flow observable over particle and antiparticle as:
\begin{equation}
    v_n^{h,{\rm 1POI}}\{2\}=\frac{\mean{V_n^h \left(V_n^{\rm ch}\right)^*}}{v_n^{\rm ch}\{2\}}=\frac{\mean{v_n^h v_n^{\rm ch} \cos [n\left(\Psi_n^h-\Psi_n^{\rm ch}\right)]}}{v_n^{\rm ch}\{2\}}.
\end{equation}
where 1POI suppresses BSQ fluctuations (see Appendix~A). 
However, two POI (2POI) observables are possible in the high luminosity era of the LHC, so we can also define:
\begin{align}
    v_n^{h}\{2\}&\equiv v_n^{h,{\rm 2POI}}\{2\}=\sqrt{\mean{V_n^{h} \left(V_n^{h}\right)^*}}=\sqrt{\mean{|V_n^{h}|^2}}
    \label{eqn:2POI}, \\
    v_n^{\pm}\{2\}&\equiv v_n^{\pm,{\rm 2POI}}\{2\}=\sqrt{\mean{|V_n^{\pm}|^2}}.
    \label{eqn:2POIpm}
\end{align}
These expressions do not depend on any mismatch between event-plane angles, since otherwise this would artificially suppress flow of PID if these angles were misaligned. 
Given our finding in Fig.\ \ref{fig:BSQ_effect}, using 2POI may lead to significantly different results because we expect the event-plane angles to be misaligned on an event-by-event basis in the presence of \iccing{}.

Since all charged flow is composed of individual hadrons, we can break up the flow into its particle ($+$) and antiparticle ($-$) contributions:
\begin{equation}
    V_n^{\rm ch} = \alpha^+V_n^+ + \alpha^-V_n^-,
    \label{eq:flow_vec_split}
\end{equation}
Using Eqs. \eqref{eq:flow_id} and \eqref{eq:flow_vec_split}, and computing the 2-particle cumulant from  Eq. \eqref{eq:vn2}, one obtains
\begin{equation}
    v_n^{\rm ch}\{2\} \approx v_n^{\varepsilon}\{2\} + \frac{1}{8}\frac{\mean{|\delta V_n^++\delta V_n^-|^2}}{v_n^{\varepsilon}\{2\}} \approx v_n^{\varepsilon}\{2\},
    \label{eq:vn_ch}
\end{equation}
as we derive in Appendix B.
In the above equation, $\mean{\delta V_n^\pm}\rightarrow 0$ when a large amount of events and PID (all charged particles include pions, protons, kaons, etc.) are taken into account, as it represents a fluctuation that is independent to $V^{\varepsilon}_n$.  
Additionally, $\mean{\delta V_n^\pm}$ is a tiny effect compared to the large contribution from $v_n^{\varepsilon}\{2\}$ such that the ratio (that is even further suppressed by a factor of $1/8$) is negligible. 

Thus, models without BSQ fluctuations can describe $v_n^{\rm ch}\{2\}$ well, since $v_n^{\rm ch}\{2\}$ predominately comes from anisotropies in the initial energy density, when averaged over many events.

Returning to the number of POI, one can see that 1POI would still wash out the effects of BSQ charge fluctuations, which has previously been shown in \cite{Plumberg:2024leb} for PID flow.
The main reason that BSQ fluctuation effects are washed out for 1POI is because $\delta V_n$ is uncorrelated with $V_n^{\rm ch}$ such that we cannot isolate $\delta V_n$ directly and linear terms like $\langle \delta V_n^{\pm}V_n^{\varepsilon}\rangle \sim 0$ vanish. 
However, if  we explore individual anisotropic flow for particles and antiparticles using the 2POI method, we maximize BSQ effects. Using Eq. \eqref{eq:flow_id} in \eqref{eqn:2POI} one has:
\begin{equation}
    v_n^\pm\{2\} \approx v_n^{\varepsilon}\{2\} + \frac{1}{2}\frac{\mean{|\delta V_n^\pm|^2}}{v_n^{\varepsilon}\{2\}},  
\label{eq:vn2_id}
\end{equation}
where the impact of BSQ in $v_n^\pm\{2\}$ is of second order in $\delta V_n^\pm$ (see derivation  in Appendix B).

\begin{figure}[ht!]
   \centering
   \includegraphics[width=0.99\columnwidth]{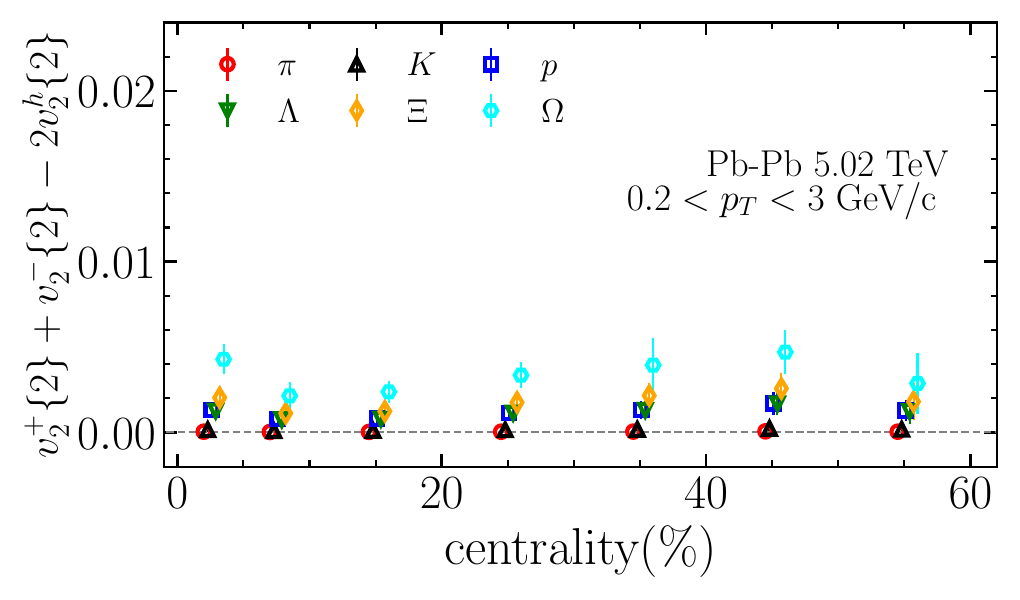}
   \caption{Relation between the elliptic flow from charged and anti-charged with the its identified hadron, relation (\ref{eq:constrain}), using \iccing{}+\ccake{} simulations for Pb-Pb collisions at $\sqrt{s_{NN}}=5.02$ TeV and $0$--$5\%$ centrality, with the same $p_T$ kinematic cuts used by the ALICE Collaboration. Identified particles were offset along the horizontal axis for clarity.}
   \label{fig:condition}
\end{figure}

{\it Results.}––To validate the hypothesis of linear fluctuation in the anisotropic flow caused by the initial energy density profile, as described in Eq.\ (\ref{eq:flow_id}), and simultaneously determine the magnitude of the fluctuation, one can combine anisotropic flow for a given species with particle and antiparticle. Using our framework for separating the $\varepsilon$ contribution from BSQ fluctuations, we isolate just the contribution of the flow magnitude fluctuations
\begin{equation}
    v_n^+\{2\}+v_n^-\{2\}-2v_n^h\{2\} \approx \frac{\mean{|\delta V_n^+|^2} + \mean{|\delta V_n^-|^2}}{4 v_n^{\varepsilon}\{2\}}\ge 0.
    \label{eq:constrain}
\end{equation}
Thus, the sum of the separate particle and antiparticle $ v_n^+\{2\}+v_n^-\{2\}$ is greater than the $v_n$ from combining both particle and antiparticles into the identified charge hadron $v_n^h\{2\}$ by Eq. \eqref{eqn:2POI}. 
In Fig \ref{fig:condition} we show the flow magnitude fluctuations results for $n=2$ versus centrality for $\pi$, $K$, $p$, $\Lambda$, $\Xi$, and $\Omega$. The flow magnitude fluctuations are extremely small ($\lesssim$1\%) and demonstrate almost no variation across centrality. The results for $n=3$ are similar (not shown).

Next, we try instead to take the ratio of either the particle or antiparticle $v_n^\pm\left\{2\right\}$ to the combined $v_n^h\left\{2\right\}$. 
For no BSQ charge fluctuations, the ratio must be exactly unity; otherwise we expect a small (positive) deviation from 1:
 \begin{equation}\label{eqn:ratio_vnCHG}
     \frac{v_n^\pm\left\{2\right\}}{v_n^h\left\{2\right\}}\approx 1+ \frac{1}{2}\frac{\mean{|\delta V_n^\pm|^2}}{\mean{|V_n^{h}|^2}},
 \end{equation}
 where we expect a more significant effect since the BSQ charge fluctuations and the $h$ contributions are quadratic.

 \begin{figure}[ht!]
    \centering
    \includegraphics[width=0.99\columnwidth]{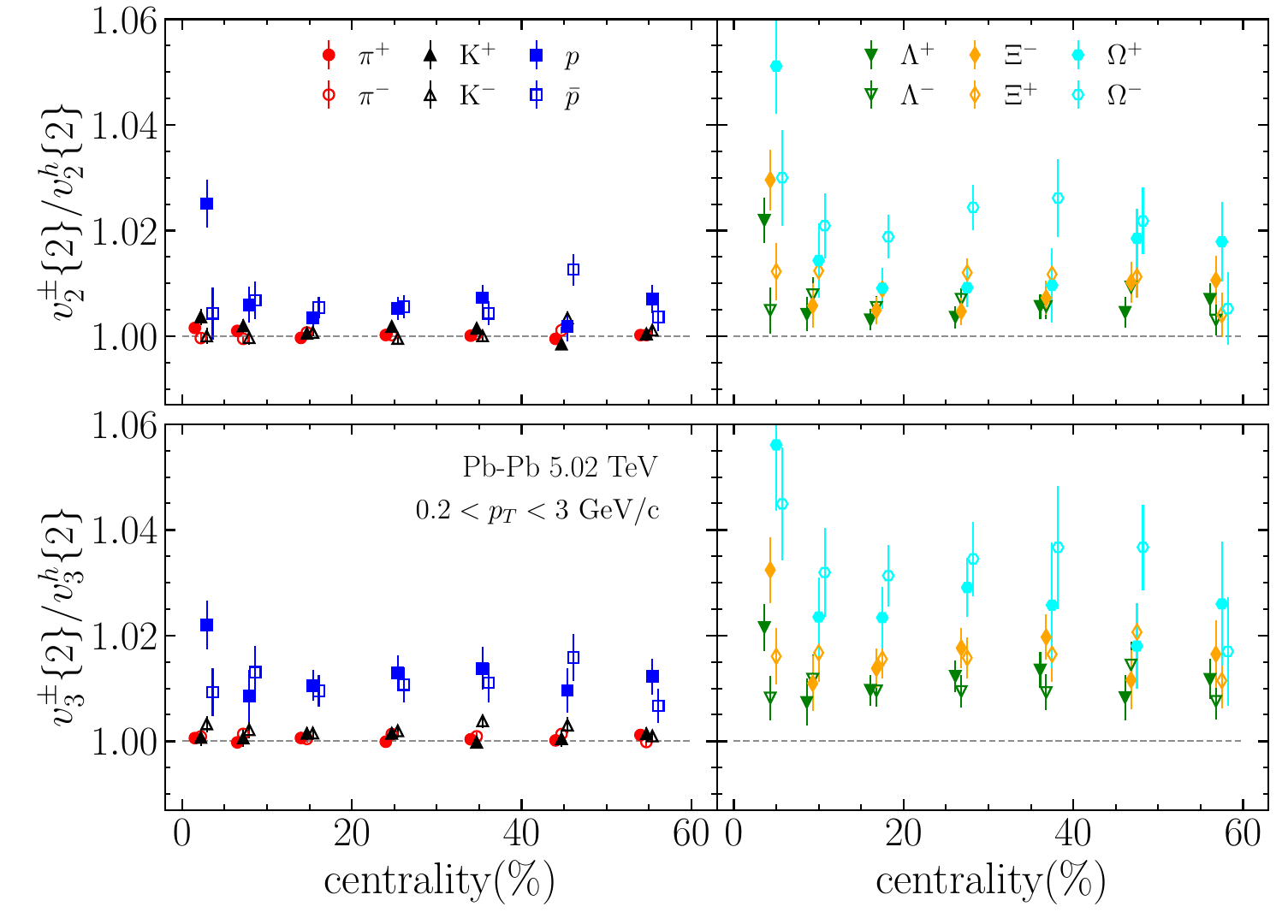}
    \caption{(Top row) Ratio between the elliptic flow of (anti)particles and its identified hadron (using both particle and antiparticle, $v_2^h$), for several species. (Bottom row) The same, with the triangular flow. Simulations using \iccing{}+\ccake{}. A ratio different from one indicates BSQ effects.
    }
    \label{fig:v_ratio}
\end{figure}
 
 In Fig. \ref{fig:v_ratio}  we calculate the ratio from Eq.~(\ref{eqn:ratio_vnCHG}) separating the previously mentioned PID into their separate particle/antiparticle contributions versus centrality. 
 We find that the size of the effect is strongly dependent on the type of particle. 
 More specifically, the size of the BSQ fluctuations is strongest for heavier particles that carry multiple conserved charges. 
Generally, these heavy particles are suppressed due to their masses at vanishing densities, but large fluctuations in chemical potentials allow for them to be produced more easily in certain regimes of the fluid.
 Pions (Q) and kaons (S, Q) carry 1-2 conserved charges each and are very light such that they have almost no visible effect. 
 Protons (B, Q) and Lambdas (B, S) are heavier, carrying two conserved charges each and see a tiny deviation from unity. 
 However, the multi-strange hyperons are the most strongly affected. 
 Heavier cascades carry (B, S, Q) and see fluctuations on the order of $\sim$1--2\% but most interesting are the $\Omega$'s who see fluctuations up to 2.5--5\%. 
 For all particles we see a larger influence in the BSQ fluctuations for $v_3$ than $v_2$. 
Experimental results for this observable would offer valuable insight into $q\bar{q}$ production. Nevertheless, we anticipate that the signal in this particular measurement will be small and mostly relevant for heavier particles that have lower statistics.

So far the observables have only shown a sensitivity to BSQ fluctuations signal at the percent level, if we only consider 2-particle (both POI) correlations. 
Next we consider the possibility of up to four POI (4POI) and find a much stronger sensitivity to BSQ fluctuations. 
A 4-particle correlation increases the number of contributions in $|\delta V_n|^2$, enhancing the sensitivity to BSQ fluctuations. 
We can study correlations  between the magnitudes of the flow vectors originating from $+$ and $-$ particles:
\begin{equation}
f_n(+,-) \equiv \frac{\mean{|V^+_n|^2|V^-_n|^2}}{\mean{|V^-_n|^4}}
=\frac{\langle \left(v_n^{+}\right)^2 \left(v_n^{-}\right)^2\rangle }{\langle\left(v_n^{-}\right)^4\rangle },
\label{eq:f}
\end{equation}
where the denominator was arbitrarily chosen to be antiparticles without loss of generality.
In the absence of BSQ fluctuations $V_n^+=V_n^-$ and $f_n=1$. 
In the presence of BSQ fluctuations, then we expect this correlator to be $f_n<1$ because $V_n^+\neq V_n^-$, which suppresses the numerator. 
If the physics driving the magnitudes of $V_n^+$ and $V_n^-$ on an event-by-event basis had absolutely no correlation whatsoever, then $f_n=0$. 
Thus, we can think of the BSQ fluctuations as something that breaks charge to anti-charge symmetry for flow on an event-by-event basis. 

\begin{figure}[ht]
    \centering
    \includegraphics[width=0.99\columnwidth]{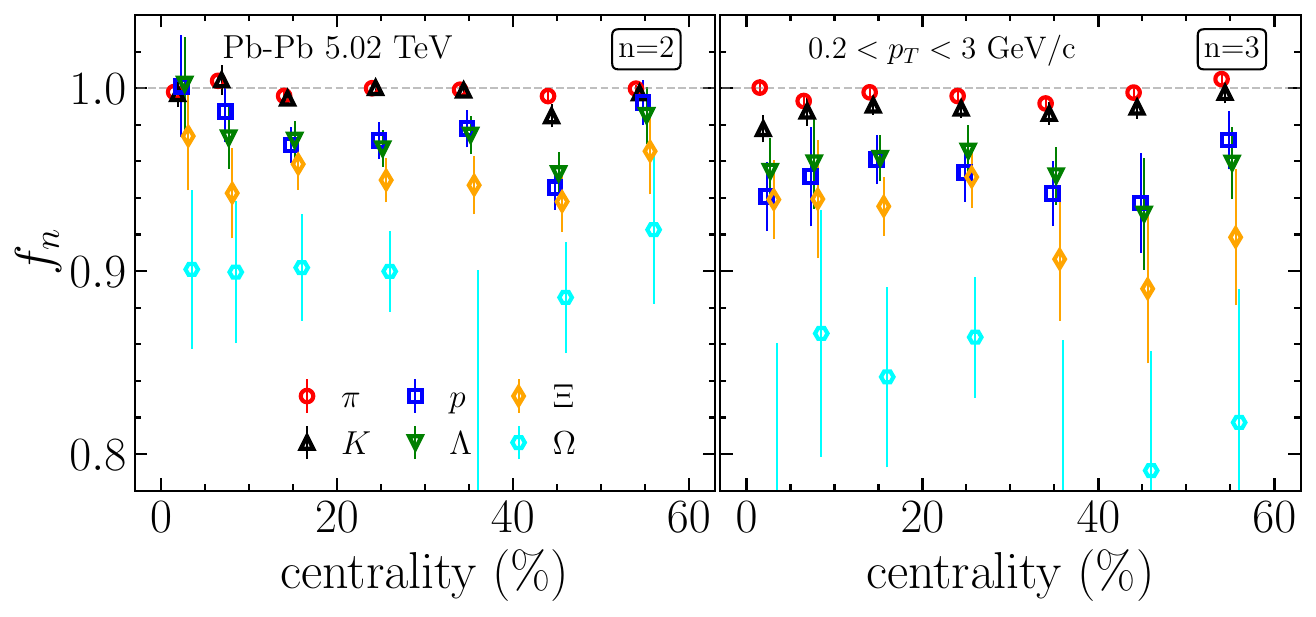}
    \caption{(Left) Correlation between the anisotropic flow of particle and antiparticle, Eq. (\ref{eq:f}), for different species. Simulations using \iccing{}+\ccake{}. A ratio different from one indicates BSQ effects.}
    \label{fig:f}
\end{figure}

The 4-particle correlator of Eq.~(\ref{eq:f}) is shown in Fig \ref{fig:f}, for elliptic (left) and triangular flow (right) for $\pi$, $K$, $p$, $\Lambda$, $\Xi$, and $\Omega$.
We find more significant deviations from unity. 
As before, pions and kaons do not see a significant effect, but protons and Lambdas appear to have up to a 2--6\% effect, $\Xi$'s around a 5--10\% effect, and $\Omega$'s up to a 10--20\% effect (strongest for $v_3$). 
However, since this signal requires 4POI it would be very challenging to measure experimentally for $\Omega$'s where we expect the largest signal. 

Ideally, we would like an observable that is sensitive to BSQ fluctuations that requires fewer POI.  
We also know that the event-plane angles for charge and anti-charge are decorrelated (see Fig.\ \ref{fig:BSQ_effect}).
Thus, we propose an event-plane correlation between $\pm$:
\begin{equation}
\mean{\cos(n\Delta\Psi_n)}\equiv
\frac{\mean{v_n^+v_n^{-} \cos[n(\Psi_n^{+}-\Psi_n^{-})]}}{\mean{(v_n^{-})^2}}=\frac{\mean{V_n^+V_n^{-}}}{\mean{|V_n^{-}|^2}},
\label{eq:eventplane}
\end{equation}
where the numerator requires a correlation between a particle and its antiparticle.  The dominator is arbitrarily chosen as ($-$) but could be just as easily ($+$), since we have already shown they provide equivalent results. 

\begin{figure}[ht]
    \centering
    \includegraphics[width=0.99\columnwidth]{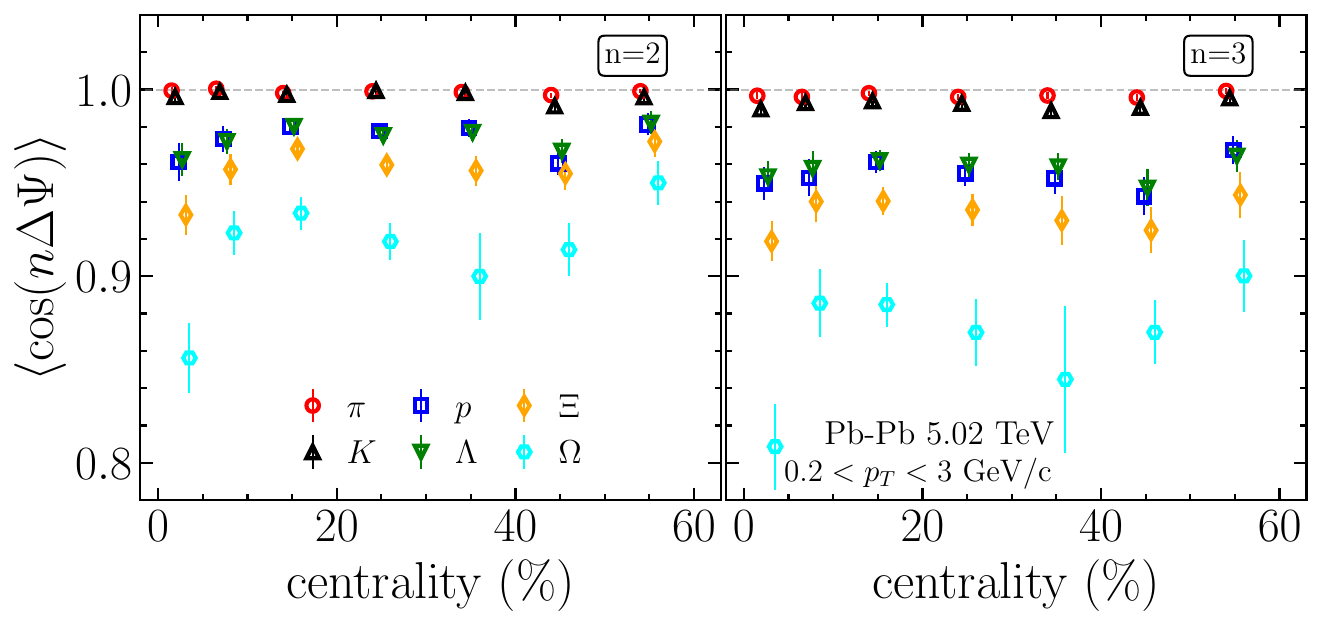}
    \caption{Event-plane correlation between particle and antiparticle for different species, Eq. (\ref{eq:eventplane}). Simulations using \iccing{}+\ccake{}. A ratio different from one indicates BSQ effects.}
    \label{fig:r}
\end{figure}

In Fig.~\ref{fig:r} we see the event-plane charge correlator versus centrality for $n=2$, $3$.
We already see up to a 2--6\% effect for protons and Lambdas, but see large effects for $\Xi$'s and $\Omega$'s where the largest effect appears in central collisions. 
Considering the event-plane correlation requires just 2POI for the numerator and denominator,  it is the ideal measurement of BSQ fluctuations. 

{\it Conclusions and Outlook.}––In this Letter, we propose new observables that rely on 2-4 POI of identified hadrons that allow us to quantify the effect of BSQ charge fluctuations that arise from $g\to q\bar{q}$ splittings in the initial state at the LHC. 
We make the first predictions for these observables and suggest that one could measure up to a $5\%$ effect for 2POI flow, Eqs. \eqref{eq:constrain} and \eqref{eqn:ratio_vnCHG}, and larger effects using correlations between particle and antiparticle, Eqs. \eqref{eq:f} and \eqref{eq:eventplane} where the largest effects are seen in central collisions. In fact, up to a $10$--$20\%$ effect could be observed if $\Omega$ baryons were possible to measure with 2+ POI and statistical significance. 
However, even using lighter particles like protons may have up to a $6\%$ effect that may be possible to measure as well. 
We note that we take ratios of flow observables such that correlated errors cancel. 
Additionally, we have designed the observables to be either 0 or 1 in the absence of fluctuations, making finding a signal easier. 

Our focus has been on BSQ fluctuations, but these observables may be of interest to studies of the chiral magnetic effect  (CME) \cite{Fukushima:2008xe,ALICE:2017sss,CMS:2017lrw,STAR:2021mii} that also should lead to non-trivial splitting of electric charge. 
In fact, our results here could be used as a background for the CME, which will be studied in the future. 
Moreover, the type of observables developed here could be used to explore other conserved charges like charm using a framework like \cite{Capellino:2022nvf,Capellino:2023cxe}.
Finally, it would be interesting to study these observables at finite baryon densities as well since the role of the equation of state (especially if a critical point exists) and of BSQ diffusion, may play an interesting role. 
As a final comment, we have made the simple assumption here that transport coefficients are constant across chemical potentials. 
While we anticipate a larger effect would be seen if we considered transport coefficients that changed across the phase diagram, further systematic analysis is required to quantify the impact of transport coefficients on these observables. This will be explored in future works.

{\it Acknowledgments.}––J.N.H., J.S.S.M, D.A. acknowledge support from the US-DOE Nuclear Science Grant No. DE-SC0023861 and within the framework of the Saturated Glue (SURGE) Topical Theory Collaboration. 
Several of the authors acknowledge the hospitality of the Kavli Institute for Theoretical Physics during “The Many
Faces of Relativistic Fluid Dynamics” Program, while this manuscript was being completed, which is supported by the National Science Foundation (NSF)  under Grant No. NSF PHY-1748958.  
This work was also supported in part by the NSF under grant number OAC-2103680 within the MUSES Collaboration.
F.G.G.~is supported by CNPq (Conselho Nacional de Desenvolvimento Cientifico) through 306762/2021-8 and by the Fulbright Program, is partially supported by CNPq through the INCT-FNA grant 312932/2018-9 and the Universal Grant 409029/2021-1.
Calculations were performed on the Illinois Campus Cluster, a computing resource that is operated by the Illinois Campus Cluster Program (ICCP) in conjunction with the National Center for Supercomputing Applications (NCSA), and which is supported by funds from the University of Illinois at Urbana-Champaign.  

\bibliography{references}

\appendix
\section{Appendix A: Why 2POI instead 1POI}

In order to understand why 1POI is not sufficient to access BSQ charges, we need to use the flow from 1POI. Assuming $V_n^\prime$ is from a given PID either particles only,  antiparticles only, or particle+antiparticle (but not all charged particles), one has
\begin{align}
    v_n^{\rm 1POI}\{2\} =\frac{\mean{V_n^{\prime}V_n^{\rm ch*}}}{\sqrt{\mean{|V_n^{\rm ch*}|^2}}}&\approx\frac{\mean{V_n^{\varepsilon}V_n^{\rm 
 ch*}}+\mean{\delta V_n^{\prime}V_n^{\rm ch}}}{v_n^{\rm ch}\{2\}} \nonumber \\
    &\approx \frac{\mean{V_n^{\varepsilon}V_n^{\rm ch*}}}{v_n^{\rm ch}\{2\}}.
\end{align}
As the flow for all charged particles is not correlated with the POI fluctuations, thus $\mean{\delta V_n^{\prime} V_n^{\rm ch*}} = \mean{\delta V_n^{\prime}}\mean{V_n^{\rm ch*}}\approx 0$, i.e., there is no first order BSQ terms in 1POI. However, in the case of 2POI, there is a small BSQ correction (see Eq. \eqref{eq:A_v_PID} in Appendix B).

\section{Appendix B: 2POI and the influence of BSQ fluctuations}

Here we detail our derivation of the breakdown of the anisotropic flow observables into their $\varepsilon$ components and that which is sensitive to BSQ fluctuations. 

 A generic 2-particle correlation anisotropic flow $v_n\{2\}$ is computed from the flow vector (assuming that one is always calculating the same particles as in both particles are all charged particles or both are POI) as
\begin{equation}
    V^\prime_n = \frac{\sum^\prime e^{in\phi}}{M}=v_n^\prime e^{in\Psi_n^\prime},
\end{equation}
where $v_n^\prime$ is the anisotropic flow and $\Psi_m^\prime$ is the event plane angle for an event. Thus,
\begin{equation}
  v_n^\prime\{2\}\equiv \sqrt{\meansq{v_n^\prime}} = \sqrt{\mean{|V_n^\prime|^2}}.
\end{equation}
Here, $V_n^\prime$ can be the case where the sum is made upon a given PID either particles only $V_n^+$,  antiparticles only $V_n^-$, particle+antiparticle $V_n^h$, and for all charged particles $V_n^{\rm ch}$. Due to BSQ charges, the flow vector has two components
\begin{equation}
    V_n^\prime = V_n^{\varepsilon} + \delta V_n^\prime,
\end{equation}
and can be used to obtain $v_n\{2\}$,
\begin{equation}
    \mean{|V_n^\prime|^2} = \mean{|V_n^{\varepsilon}|^2} + 2\Re(\mean{V_n^{\varepsilon}\delta {V_n^{\prime *}}}) + \mean{|\delta V_n^\prime|^2}.  
\end{equation}
As $\delta V_n^\prime$ represent a small perturbation coming from the fluctuations of BSQ charges, the average correlation between it and the main contribution (energy density $V_n^{\varepsilon}$), returns $\mean{V_n^{\varepsilon}\delta V_n^{\prime*}}\approx 0$, since the fluctuations are not correlated with the background. Therefore, since $V_n^{\varepsilon}\gg \delta V_n^\prime$, and in the case of particle/antiparticle
\begin{align}
v_n^\prime\{2\}&=\sqrt{\mean{|V_n^\prime|^2}}\approx \sqrt{\mean{|V_n^{\varepsilon}|^2}}\left(1+\frac{\mean{|\delta V_n^\prime|^2}}{\mean{|V_n^{\varepsilon}|^2}}\right)^{\frac{1}{2}}\nonumber \\
&=v_n^{\varepsilon}\{2\}\left(1+\frac{\mean{|\delta V_n^\prime|^2}}{v_n^{\varepsilon}\{2\}^2}\right)^\frac{1}{2}\approx v_n^{\varepsilon}\{2\}+\frac{1}{2}\frac{\mean{|\delta V_n^\prime|^2}}{v_n^{\varepsilon}\{2\}}.
\label{eq:A_vn}
\end{align}
Only the first non-vanishing term was kept, since the next correction is  negligible.  The above expression holds for particle ($+$), antiparticle ($-$), particle+antiparticle ($h$) and all charged hadrons (ch). 

For charged particles, using Eq. \eqref{eq:flow_vec_split} in \eqref{eq:A_vn}, and $\delta V_n^{\prime}\equiv\delta V_n^{\rm ch} = (\delta V_n^+ + \delta V_n^-)/2$, we have Eq. \eqref{eq:vn_ch}
\begin{equation}
    v_n^{\rm ch}\{2\} \approx v_n^{\varepsilon}\{2\} + \frac{1}{8}\frac{\mean{|\delta V_n^++\delta V_n^-|^2}}{v_n^{\varepsilon}\{2\}}.
\end{equation}
For PID, one has $\delta V_n^\prime = \delta V_n^{\pm}$, then Eq.~\eqref{eq:vn2_id} becomes
\begin{equation}
    v_n^\pm\{2\} \approx v_n^{\varepsilon}\{2\} + \frac{1}{2}\frac{\mean{|\delta V_n^\pm|^2}}{v_n^{\varepsilon}\{2\}}.
    \label{eq:A_v_PID}
\end{equation}

Dividing Eq.~\eqref{eq:A_v_PID} by $v_n^{\varepsilon}\{2\}$, which is in first order $v_n^{\varepsilon}\{2\}\approx v_n^{\rm h}\{2\}=\sqrt{\mean{|V_n^h|}}$, one obtains Eq. \eqref{eqn:ratio_vnCHG}.
This clarifies the reason for the tiny impact of BSQ observed when employing the anisotropic flow from an identified hadron.  It is essential to correlate two or more flow vectors from particles and antiparticles to enhance contributions to the observable.

\end{document}